\documentclass[twocolumn,notitlepage,nofootinbib]{revtex4-2}
\usepackage{mathtools,amssymb,bm,graphicx,hyperref}
\allowdisplaybreaks[1]

\newcommand\+{\dagger}
\renewcommand\*{\star}
\newcommand\<{\langle}
\renewcommand\>{\rangle}
\newcommand\0{{\bm{0}}}
\newcommand\A{{\bm{A}}}
\newcommand\E{{\bm{E}}}
\newcommand\p{{\bm{p}}}
\newcommand\q{{\bm{q}}}
\newcommand\x{{\bm{x}}}
\renewcommand\O{\mathcal{O}}
\renewcommand\in{\mathrm{in}}
\newcommand\out{\mathrm{out}}
\newcommand\vac{\mathrm{vac}}
\newcommand\T{\mathrm{T}}
\DeclareMathOperator\diag{diag}
\DeclareMathOperator\tr{tr}
\DeclareMathOperator\Tr{Tr}

\begin{document}

\title{Full counting statistics of Schwinger pair production and annihilation}

\author{Yusuke Nishida}
\affiliation{Department of Physics, Tokyo Institute of Technology,
Ookayama, Meguro, Tokyo 152-8551, Japan}

\date{June 2021}

\begin{abstract}
We study the probability distribution of the number of particle and antiparticle pairs produced via the Schwinger effect when a uniform but time-dependent electric field is applied to noninteracting scalars or spinors initially at a thermodynamic equilibrium.
We derive the formula for the characteristic function by employing techniques in mesoscopic physics, reflecting a close analogy between the Schwinger effect and mesoscopic tunneling transports.
In particular, we find that the pair production in a medium is enhanced (suppressed) for scalars (spinors) due to the Bose stimulation (Pauli blocking).
Furthermore, in addition to the production of accelerated pairs by the electric field, the annihilation of decelerated pairs is found to take place in a medium.
Our formula allows us to extract the probability distributions in various situations, such as those obeying the generalized trinomial statistics for spin-momentum resolved counting and the bidirectional Poisson statistics for spin-momentum unresolved counting.
\end{abstract}

\maketitle

\section{Introduction}
Together with earlier suggestions by Sauter, Heisenberg, and Euler~\cite{Sauter:1931,Heisenberg:1936}, Schwinger in 1951 predicted that pairs of electron and positron are produced out of a vacuum subjected to an external electric field~\cite{Schwinger:1951}.
This phenomenon is now known as the Schwinger effect, which is a manifestation of the fact that the quantum vacuum is no longer a classical empty space but undergoes virtual pair excitations, being the Copernican revolution of our view on the vacuum.

Although the Schwinger effect constitutes one of the most significant predictions of quantum electrodynamics, its experimental observation has been elusive because the mean number of produced pairs is exponentially suppressed if the electric field is below the critical strength set by the electron mass.
However, new prospects to reach and exceed the critical electric field have recently emerged in high-intensity laser facilities~\cite{Piazza:2012}, as well as in relativistic heavy-ion collisions~\cite{Baur:2007}, which have reaccelerated research into the Schwinger effect calling for further insights as a realistic possibility~\cite{Gelis:2016}.

The purpose of this paper is to study the probability distribution of the number of produced pairs, going beyond just the mean value usually considered.
This is partly motivated by an estimation that the mean number of produced pairs is at best far from macroscopic~\cite{Ringwald:2001,Alkofer:2001}, so that its event-by-event fluctuations should be important.
It is also evident that the probability distribution brings out much more information from the system, as concisely phrased by ``the noise is the signal''~\cite{Landauer:1998}.
In fact, such a probability distribution of the number of transmitted charges is referred to as full counting statistics and has been one of the main research streams in mesoscopic physics~\cite{Nazarov,Nazarov-Blanter}.
Because the Schwinger effect can also be viewed as a quantum tunneling phenomenon~\cite{Holstein:1999}, common techniques can be employed to derive the formula for full counting statistics, which reflects a close analogy between the Schwinger effect and mesoscopic tunneling transports.

\section{Schwinger pair production}
We consider noninteracting scalars ($S=0$) or spinors ($S=1/2$) with mass $m$ and charge $q$ subjected to an external electric field, which obey the Klein-Gordon or Dirac equation, respectively.
The time-dependent electric field is assumed to be spatially uniform and turned off in the infinite past and future, so that the gauge is chosen to be $\A(t)=\int_t^\infty\!dt'\E(t')$.
We will set $\hbar=c=k=1$ throughout this paper.

One convenient way to describe the Schwinger effect is based on the canonical quantization of the single-particle field operator $\hat\Psi(x)$~\cite{Rumpf:1976a,Rumpf:1976b,Rumpf:1978}.
Because its detailed accounts are readily available, for example, in Refs.~\cite{Tanji:2009,Fukushima:2009,Gelis:2016}, we only outline the basic ingredients needed for our following discussions.
The field operator can be expanded on a basis of eigenfunctions of the Klein-Gordon or Dirac equation as
\begin{align}
\hat\Psi(x) = \sum_{s,\p}\left[\hat{a}_{s,\p}^\in\psi_{s,\p}^{\in+}(x)
+ \hat{c}_{-s,-\p}^{\in\+}\psi_{s,\p}^{\in-}(x)\right],
\end{align}
where $s=-S,-S+1,\dots,S$ and $\p$ label spin and canonical momentum, respectively, assuming the system in a periodic box.
By choosing the eigenfunctions so as to satisfy $\lim_{t\to-\infty}\psi_{s,\p}^{\in\pm}(x)\sim e^{i\p\cdot\x\mp i\sqrt{m^2+[\p-q\A(-\infty)]^2}\,t}$, $\hat{a}_{s,\p}^\in$ and $\hat{c}_{-s,-\p}^\in$ are identified as annihilation operators of particle and antiparticle in the infinite past.
On the other hand, the same field operator can also be expanded on a different basis as
\begin{align}
\hat\Psi(x) = \sum_{s,\p}\left[\hat{a}_{s,\p}^\out\psi_{s,\p}^{\out+}(x)
+ \hat{c}_{-s,-\p}^{\out\+}\psi_{s,\p}^{\out-}(x)\right],
\end{align}
where $\lim_{t\to+\infty}\psi_{s,\p}^{\out\pm}(x)\sim e^{i\p\cdot\x\mp i\sqrt{m^2+\p^2}\,t}$ so that $\hat{a}_{s,\p}^\out$ and $\hat{c}_{-s,-\p}^\out$ are in turn annihilation operators of particle and antiparticle in the infinite future.
We note that, because of $\A(+\infty)=\0$ in our gauge choice, $\p$ coincides with the kinetic momentum in the infinite future, which is accelerated from $\p-q\A(-\infty)$ being the kinetic momentum in the infinite past.

Because each of $\psi_{s,\p}^{\in\pm}(x)$ is a superposition of $\psi_{s,\p}^{\out+}(x)$ and $\psi_{s,\p}^{\out-}(x)$, $\hat{a}_{s,\p}^\out$ and $\hat{c}_{-s,-\p}^{\out\+}$ can be expressed in terms of $\hat{a}_{s,\p}^\in$ and $\hat{c}_{-s,-\p}^{\in\+}$ as
\begin{align}\label{eq:bogoliubov}
\begin{pmatrix}
\hat{a}_{s,\p}^\out \\ \hat{c}_{-s,-\p}^{\out\+}
\end{pmatrix}
= U_{s,\p}
\begin{pmatrix}
\hat{a}_{s,\p}^\in \\ \hat{c}_{-s,-\p}^{\in\+}
\end{pmatrix}.
\end{align}
Here it is assumed that the spin basis is chosen so that different $s$ sectors are decoupled.
In order to preserve the commutation (anticommutation) relations for bosons (fermions) with $\lambda=(-1)^{2S}$,
\begin{subequations}
\begin{align}
& [\hat{a}_{s,\p},\hat{a}_{s,\p}^\+]_\lambda
\equiv \hat{a}_{s,\p}\hat{a}_{s,\p}^\+ - \lambda\hat{a}_{s,\p}^\+\hat{a}_{s,\p} = 1, \\
& [\hat{c}_{-s,-\p},\hat{c}_{-s,-\p}^\+]_\lambda = 1, \qquad \text{others} = 0,
\end{align}
\end{subequations}
$U_{s,\p}$ must be a $2\times2$ paraunitary (unitary) matrix satisfying
\begin{align}\label{eq:matrix}
U_{s,\p}
\begin{pmatrix}
1 & 0 \\ 0 & -\lambda
\end{pmatrix}
U_{s,\p}^\+
=
\begin{pmatrix}
1 & 0 \\ 0 & -\lambda
\end{pmatrix}.
\end{align}
Its general form is provided by
\begin{align}
U_{s,\p} = e^{i\delta_{s,\p}}
\begin{pmatrix}
\alpha_{s,\p} & \gamma_{s,\p} \\
\lambda\gamma_{s,\p}^* & \alpha_{s,\p}^*
\end{pmatrix}
\end{align}
with $|\alpha_{s,\p}|^2-\lambda|\gamma_{s,\p}|^2=1$, which constitutes the Bogoliubov transformation for bosons (fermions).

Suppose that the system is initially in the vacuum, $\hat{a}_{s,\p}^\in|\vac\>=\hat{c}_{-s,-\p}^\in|\vac\>=0$, with no particles and antiparticles present.
After the electric field is applied, we however find that pairs of particle and antiparticle are produced according to $\<\vac|\hat{a}_{s,\p}^{\out\+}\hat{a}_{s,\p}^\out|\vac\>=\<\vac|\hat{c}_{-s,-\p}^{\out\+}\hat{c}_{-s,-\p}^\out|\vac\>=|\gamma_{s,\p}|^2$.
This is none other than the Schwinger effect.
The mean number of produced pairs with spins and momenta of $\pm(s,\p)$ for particle (upper sign) and antiparticle (lower sign) is provided by the Bogoliubov coefficient $|\gamma_{s,\p}|^2$, which is to be determined by solving the Klein-Gordon or Dirac equation for a given electric field.
Such analyses were carried out for a variety of solvable temporal profiles~\cite{Breev:arxiv}, including the constant pulse $\E(t)=\E_0\,\theta(T/2-|t|)$ and the Sauter pulse $\E(t)=\E_0/\cosh^2(2t/T)$~\cite{Tanji:2009,Fukushima:2009,Gelis:2016}.
In either case, $T$ is the duration of the applied electric field with $q\A(-\infty)=q\E_0T$ and the long-time limit leads to
\begin{align}\label{eq:coefficient}
\lim_{T\to\infty}|\gamma_{s,\p}|^2 = \theta(p_\parallel)\,\theta(qE_0T-p_\parallel)
\exp\!\left(-\pi\frac{m^2+\p_\perp^2}{qE_0}\right)
\end{align}
for $|p_\parallel|,|p_\parallel-qE_0T|\gg m$~\cite{Cohen:2008}, where $p_\parallel$ and $\p_\perp$ are momenta parallel and perpendicular to $q\E_0$, respectively.

\section{Characteristic function}
We now wish to determine the probability distribution of the number of produced pairs instead of just the mean value.
For the sake of generality and comparison with mesoscopic tunneling transports, we assume that the system is initially at a thermodynamic equilibrium with temperature $\beta^{-1}$ and chemical potential $\mu$.
The joint probability to produce $N_{s,\p}$ particles and $\bar N_{s,\p}$ antiparticles for each spin and momentum (i.e., $N_{s,\p}=N_{s,\p}^\out-N_{s,\p}^\in$ being the particle number in the infinite future minus that in the infinite past) is denoted by $P(\{N\},\{\bar N\})$.
It is rather convenient to consider its Fourier series,
\begin{align}
\chi(\{\theta\},\{\bar\theta\})
&\equiv \sum_{\{N\}}\sum_{\{\bar N\}}P(\{N\},\{\bar N\}) \notag\\
&\times e^{i\sum_{s,\p}(\theta_{s,\p}N_{s,\p}+\bar\theta_{s,\p}\bar N_{s,\p})},
\end{align}
defining the characteristic function.
We note that the cumulants are generated by a power series expansion of $\ln\chi(\{\theta\},\{\bar\theta\})$ and the probability distribution is recovered according to
\begin{align}\label{eq:probability}
P(\{N\},\{\bar N\}) &= \left[\prod_{s,\p}\iint_{-\pi}^\pi\!
\frac{d\theta_{s,\p}d\bar\theta_{s,\p}}{(2\pi)^2}\right]
\chi(\{\theta\},\{\bar\theta\}) \notag\\
&\times e^{-i\sum_{s,\p}(N_{s,\p}\theta_{s,\p}+\bar N_{s,\p}\bar\theta_{s,\p})}.
\end{align}

The characteristic function can be expressed in terms of the creation and annihilation operators of particle and antiparticle as
\begin{align}
\chi(\{\theta\},\{\bar\theta\}) &= \frac1Z\Tr\Bigl[e^{-\beta\hat{H}}
e^{i\sum_{s,\p}(\theta_{s,\p}\hat{a}_{s,\p}^{\out\+}\hat{a}_{s,\p}^\out
+ \bar\theta_{s,\p}\hat{c}_{s,\p}^{\out\+}\hat{c}_{s,\p}^\out)} \notag\\
&\times e^{-i\sum_{s,\p}(\theta_{s,\p}\hat{a}_{s,\p}^{\in\+}\hat{a}_{s,\p}^\in
+ \bar\theta_{s,\p}\hat{c}_{s,\p}^{\in\+}\hat{c}_{s,\p}^\in)}\Bigr].
\end{align}
Here $\hat{H}=\sum_{s,\p}[(E_\p-\mu)\hat{a}_{s,\p}^{\in\+}\hat{a}_{s,\p}^\in+(E_{-\p}+\mu)\hat{c}_{s,\p}^{\in\+}\hat{c}_{s,\p}^\in]$ is the initial Hamiltonian with the dispersion relation of
\begin{align}
E_\p = \sqrt{m^2+\left[\p-q\A(-\infty)\right]^2},
\end{align}
$Z=\Tr[e^{-\beta\hat{H}}]$ is the partition function, and the trace is taken in the Fock space.
We then introduce an operator of two components by $\hat{b}_{s,\p}=(\hat{a}_{s,\p}^\in,\hat{c}_{-s,-\p}^{\in\+})^\T$ and its conjugate by $\hat{b}_{s,\p}^\*=(\hat{a}_{s,\p}^{\in\+},-\lambda\hat{c}_{-s,-\p}^\in)$, so that their components obey the commutation (anticommutation) relations for bosons (fermions),
\begin{subequations}
\begin{align}
& [\hat{b}_{r,\p}^i,\hat{b}_{s,\q}^{\*j}]_\lambda
= \delta_{ij}\delta_{rs}\delta_{\p\q} \qquad (i,j=1,2), \\
& [\hat{b}_{r,\p}^i,\hat{b}_{s,\q}^j]_\lambda
= [\hat{b}_{r,\p}^{\*i},\hat{b}_{s,\q}^{\*j}]_\lambda = 0.
\end{align}
\end{subequations}
With the help of the Bogoliubov transformation in Eq.~(\ref{eq:bogoliubov}) and $U_{s,\p}^{-1}=\diag(1,-\lambda)\,U_{s,\p}^\+\diag(1,-\lambda)$ following from Eq.~(\ref{eq:matrix}), the characteristic function now reads
\begin{align}\label{eq:fock-space}
\chi(\{\theta\},\{\bar\theta\}) &= \prod_{s,\p}\frac1{Z_{s,\p}}
\tr_{s,\p}\Bigl[e^{-\beta\hat{b}_{s,\p}^\*H_{s,\p}\hat{b}_{s,\p}} \notag\\
&\times e^{i\hat{b}_{s,\p}^\*U_{s,\p}^{-1}\Theta_{s,\p}U_{s,\p}\hat{b}_{s,\p}}
e^{-i\hat{b}_{s,\p}^\*\Theta_{s,\p}\hat{b}_{s,\p}}\Bigr],
\end{align}
where $\Theta_{s,\p}=\diag(\theta_{s,\p},-\bar\theta_{-s,-\p})$, $H_{s,\p}=\diag(E_\p-\mu,-E_\p-\mu)$, and $Z_{s,\p}=\tr_{s,\p}[e^{-\beta\hat{b}_{s,\p}^\*H_{s,\p}\hat{b}_{s,\p}}]$ are introduced and the lowercase trace is taken in a Fock space spanned by $(\hat{b}_{s,\p}^{\*1})^n(\hat{b}_{s,\p}^{\*2})^{\bar n}|0\>$ with $\hat{b}_{s,\p}^1|0\>=\hat{b}_{s,\p}^2|0\>=0$ for the given set of $(s,\p)$.

Remarkably, the trace in the Fock space can be transformed into the determinant in the single-particle Hilbert space according to
\begin{align}
\tr\bigl[e^{\hat{O}(A)}e^{\hat{O}(B)}\cdots e^{\hat{O}(C)}\bigr]
= \det\bigl[1-\lambda\,e^Ae^B\cdots e^C\bigr]^{-\lambda}.
\end{align}
Here $\hat{O}(A)=\hat{b}^{\*\!}A\,\hat{b}$ is a bilinear operator with an arbitrary square matrix $A$ and the above identity was proven with $[\hat{O}(B),\hat{O}(C)]=\hat{O}([B,C])$ and the Baker-Campbell-Hausdorff formula by Klich~\cite{Klich:2003}.
Its application to Eq.~(\ref{eq:fock-space}) leads to $Z_{s,\p}=\det\!\left[1-\lambda e^{-\beta H_{s,\p}}\right]^{-\lambda}$ and
\begin{align}
& \chi(\{\theta\},\{\bar\theta\}) \notag\\
&= \prod_{s,\p}\det\!\left[1 - \frac\lambda{e^{\beta H_{s,\p}}-\lambda}
\left(U_{s,\p}^{-1}e^{i\Theta_{s,\p}}U_{s,\p}e^{-i\Theta_{s,\p}}-1\right)\right]^{-\lambda}.
\end{align}
Finally, it is straightforward to manipulate the $2\times2$ matrices to obtain
\begin{widetext}
\begin{align}\label{eq:characteristic}
\ln\chi(\{\theta\},\{\bar\theta\})
= -\lambda\frac{V}{(2\pi)^3}\sum_s\int\!d^3\p\,\ln\Bigl[1
&- \lambda\left\{1+\lambda f_\lambda(E_\p-\mu)\right\}
\left\{1+\lambda f_\lambda(E_\p+\mu)\right\}|\gamma_{s,\p}|^2
\bigl(e^{i\theta_{s,\p}+i\bar\theta_{-s,-\p}}-1\bigr) \notag\\
&- \lambda\,f_\lambda(E_\p-\mu)\,f_\lambda(E_\p+\mu)\,|\gamma_{s,\p}|^2
\bigl(e^{-i\theta_{s,\p}-i\bar\theta_{-s,-\p}}-1\bigr)\Bigr].
\end{align}
\end{widetext}
Here the thermodynamic limit is taken with $V$ being the volume subjected to the electric field and $f_\lambda(E_\p\mp\mu)=1/[e^{\beta(E_\p\mp\mu)}-\lambda]$ are the Bose-Einstein or Fermi-Dirac distribution functions of particles (upper sign) and antiparticles (lower sign).

The resulting formula for the characteristic function constitutes the central outcome of this paper, from which the probability distributions can be extracted in various situations as discussed below for some specific cases.
Here we make general remarks.
The term $\sim e^{i\theta_{s,\p}}$ ($e^{-i\theta_{s,\p}}$) represents a process to add (remove) one particle with spin and momentum of $(s,\p)$, whereas the term $\sim e^{i\bar\theta_{-s,-\p}}$ ($e^{-i\bar\theta_{-s,-\p}}$) represents a process to add (remove) one antiparticle with $(-s,-\p)$.
Because the characteristic function depends only on the combinations of $e^{i\theta_{s,\p}+i\bar\theta_{-s,-\p}}$ and $e^{-i\theta_{s,\p}-i\bar\theta_{-s,-\p}}$, particles and antiparticles are always produced or annihilated in pairs.

The probability for pair production per mode is quantified by
\begin{align}
F_{s,\p}^\lambda &\equiv \left[1+\lambda f_\lambda(E_\p-\mu)\right]
\left[1+\lambda f_\lambda(E_\p+\mu)\right]|\gamma_{s,\p}|^2,
\end{align}
which in the medium is enhanced for bosons due to the Bose stimulation but suppressed for fermions due to the Pauli blocking.
On the other hand, the probability for pair annihilation per mode is quantified by
\begin{align}
G_{s,\p}^\lambda \equiv f_\lambda(E_\p-\mu)\,f_\lambda(E_\p+\mu)\,|\gamma_{s,\p}|^2,
\end{align}
which is nonvanishing only in the medium because particles and antiparticles must initially be present in order for a pair to be annihilated.
According to Eq.~(\ref{eq:coefficient}), produced particles (antiparticles) tend to have final kinetic momenta in the same (opposite) direction to $q\E_0$ so as to be accelerated by the electric field, whereas annihilated particles (antiparticles) tend to have initial kinetic momenta in the opposite (same) direction to $q\E_0$ so as to be decelerated by the electric field.
The latter is actually the time-reversal process of the former, solving the same equation of motion but under different initial conditions.
Although $F_{s,\p}/G_{s,\p}=e^{2\beta E_\p}>1$ indicates that the pair production dominates over the pair annihilation on average, it does not have to be the case in individual events (see Fig.~\ref{fig:spinor}).

Our formula is indeed parallel to the Levitov formula in mesoscopic physics~\cite{Levitov:1993,Ivanov:1993}, where our particle and antiparticle degrees of freedom correspond to left and right reservoirs in mesoscopic tunneling transports~\cite{Nazarov-Blanter}.
However, the distribution functions appear differently in our characteristic function, reflecting the fact that ours involves pair production and annihilation instead of transmissions between two reservoirs.

\section{Full counting statistics}
\subsection{Spin-momentum resolved counting}
Let us first derive the probability distribution in the case where the numbers of produced particles and antiparticles are fully counted per spin and momentum.
To this end, we discretize the continuous momentum in Eq.~(\ref{eq:characteristic}) into a bunch of sufficiently small bins so that the integrand is regarded as constant in each bin.
The substitution of Eq.~(\ref{eq:characteristic}) into Eq.~(\ref{eq:probability}) then leads to
\begin{align}
P(\{N\},\{\bar N\}) = \prod_{s,\p}P_{s,\p}(N_{s,\p})\,\delta_{N_{s,\p}\bar N_{-s,-\p}},
\end{align}
where
\begin{align}
P_{s,\p}(N) &= \int_{-\pi}^\pi\frac{d\theta}{2\pi}\,e^{-iN\theta}
\bigl[1 + \lambda F_{s,\p}^\lambda + \lambda G_{s,\p}^\lambda \notag\\
&\quad - \lambda F_{s,\p}^\lambda e^{i\theta}
- \lambda G_{s,\p}^\lambda e^{-i\theta}\bigr]^{-\lambda V\Delta\p/(2\pi)^3}
\end{align}
is the marginal probability to produce $N$ particles for particular $s$ and $\p$ in a bin volume of $\Delta\p$.
We can perform the integration over $\theta$ with the help of the trinomial series to obtain
\begin{align}\label{eq:trinomial}
P_{s,\p}(N)
&= \sum_{n=0}^\infty\frac{[-\lambda V\Delta\p/(2\pi)^3]_{N+2n}}{(N+n)!\,n!} \notag\\
&\times \bigl(1 + \lambda F_{s,\p}^\lambda + \lambda G_{s,\p}^\lambda\bigr)^{-\lambda V\Delta\p/(2\pi)^3-N-2n} \notag\\
&\times \bigl(-\lambda F_{s,\p}^\lambda\bigr)^{N+n}
\bigl(-\lambda G_{s,\p}^\lambda\bigr)^n,
\end{align}
where $[z]_n\equiv z\,(z-1)\cdots(z-n+1)$ is the falling factorial.
This is, so to say, a generalized trinomial distribution with the positive (negative) exponent $-\lambda V\Delta\p/(2\pi)^3$ for fermions (bosons)~\cite{footnote1}.
Its probability distribution is extended to $N<0$ if and only if $G_{s,\p}^\lambda\neq0$, which is a consequence of the pair annihilation by applying the electric field to the medium.

On the other hand, when the electric field is applied to the vacuum with $\beta^{-1}=\mu=0$, the pair annihilation is prohibited because of $F_{s,\p}^\lambda=|\gamma_{s,\p}|^2$ and $G_{s,\p}^\lambda=0$.
The above probability distribution is then reduced to the (negative) binomial distribution for fermions (bosons) generalized to the nonintegral exponent.
In particular, its special case with $V\Delta\p/(2\pi)^3=1$ is the Bernoulli distribution for fermions or the geometric distribution for bosons~\cite{Fukushima:2009}.

\subsection{Spin-momentum unresolved counting}
Let us turn to the probability distribution in the case where only the total numbers of produced particles and antiparticles are counted without resolving their spins and momenta.
The probability to produce $N$ particles and $\bar N$ antiparticles is provided by
\begin{align}\label{eq:unresolved}
P(N,\bar N) = \iint_{-\pi}^\pi\frac{d\theta d\bar\theta}{(2\pi)^2}\,
\chi(\theta,\bar\theta)\,e^{-iN\theta-i\bar N\bar\theta},
\end{align}
where $\chi(\theta,\bar\theta)$ is obtained by setting $\theta_{s,\p}\to\theta$ and $\bar\theta_{s,\p}\to\bar\theta$ in Eq.~(\ref{eq:characteristic}).
The cumulants are generated by a power series expansion of $\ln\chi(\theta,\bar\theta)$ and can be expressed in terms of $F_{s,\p}^\lambda$ and $G_{s,\p}^\lambda$.
In particular, the mean value reads $\<N\>=\<\bar N\>=X-Y$, where
\begin{subequations}
\begin{align}
X &\equiv \frac{V}{(2\pi)^3}\sum_s\int\!d^3\p\,F_{s,\p}^\lambda, \\
Y &\equiv \frac{V}{(2\pi)^3}\sum_s\int\!d^3\p\,G_{s,\p}^\lambda
\end{align}
\end{subequations}
are constants representing the mean numbers of produced and annihilated pairs, respectively.
We note that the mean value per mode, $F_{s,\p}^\lambda-G_{s,\p}^\lambda=\left[1+\lambda f_\lambda(E_\p-\mu)+\lambda f_\lambda(E_\p+\mu)\right]|\gamma_{s,\p}|^2$, is consistent with Refs.~\cite{Hallin:1995,Kim:2007,Gavrilov:2008,Kim:2009}.

In order to proceed further without numerics, some simplification is needed and it is reasonable to assume $|\gamma_{s,\p}|^2\ll1$ because it is exponentially suppressed if the electric field is below the critical strength [see Eq.~(\ref{eq:coefficient})].
The expansion of Eq.~(\ref{eq:characteristic}) up to $\O(|\gamma_{s,\p}|^2)$ leads to
\begin{align}
\ln\chi(\theta,\bar\theta) \simeq X\bigl(e^{i\theta+i\bar\theta}-1\bigr)
+ Y\bigl(e^{-i\theta-i\bar\theta}-1\bigr),
\end{align}
which is the so-called bidirectional Poisson distribution with its $n$th cumulant provided by $X+(-1)^nY$~\cite{Levitov:2004,Esposito:2009,footnote2}.
We can then perform the integrations over $\theta,\bar\theta$ in Eq.~(\ref{eq:unresolved}) to obtain
\begin{align}\label{eq:bidirectional}
P(N,\bar N)
\simeq e^{-X-Y}\left(\frac{X}{Y}\right)^{N/2}I_{|N|}(2\sqrt{XY})\,\delta_{N\bar N},
\end{align}
where $I_n(z)$ is the modified Bessel function of the first kind.
The resulting probability distribution is extended to $N<0$ if and only if $Y\neq0$, which is again a consequence of the pair annihilation by applying the electric field to the medium.

\begin{figure}[t]
\includegraphics[width=0.9\columnwidth]{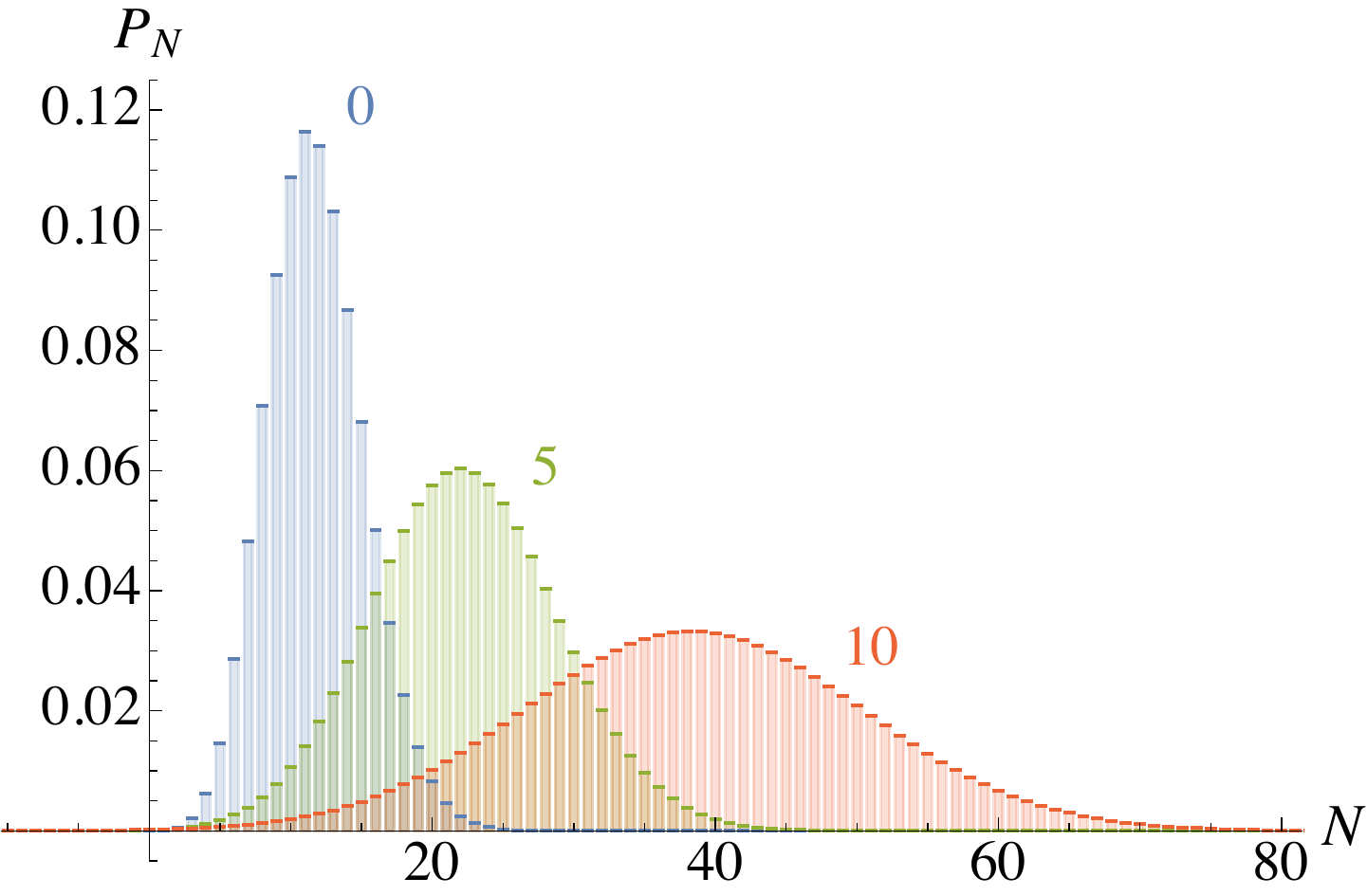}
\caption{\label{fig:scalar}
Probability distribution of the number of produced pairs for scalars at three different temperatures $\beta^{-1}/m=0$ (left), 5 (middle), 10 (right), and $\mu=0$ with the electric field fixed as described in the text.
The mean value increases as $\<N\>\approx11.8$, 22.3, 38.5 with increasing the temperature due to the Bose stimulation.}
\end{figure}

\begin{figure}[t]
\includegraphics[width=0.9\columnwidth]{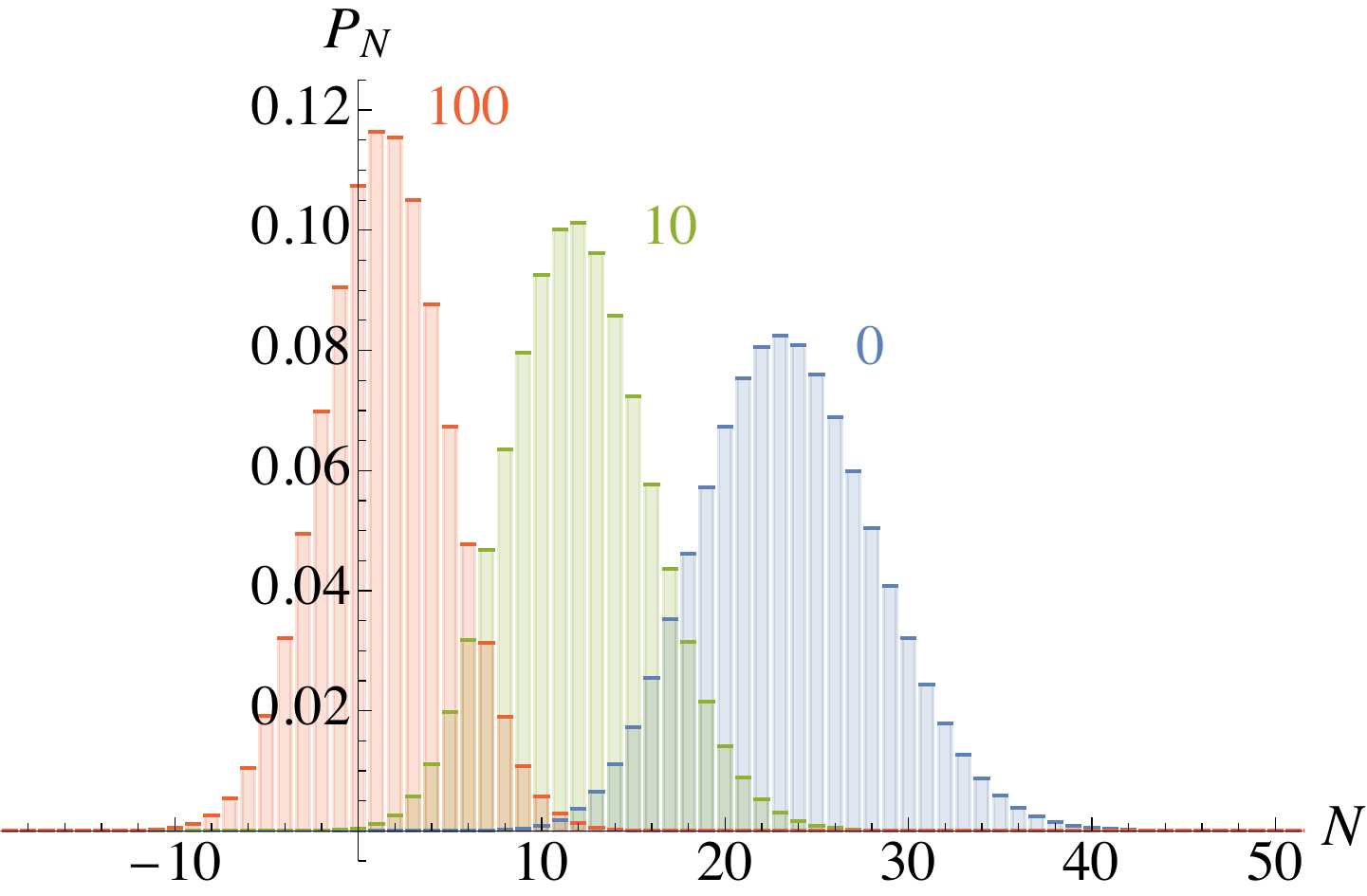}
\caption{\label{fig:spinor}
Probability distribution of the number of produced pairs for spinors at three different temperatures $\beta^{-1}/m=0$ (right), 10 (middle), 100 (left), and $\mu=0$ with the electric field fixed as described in the text.
The mean value decreases as $\<N\>\approx23.5$, 12.1, 1.48 with increasing the temperature due to the Pauli blocking.
The cumulative probability for pair annihilation reads $\sum_{N<0}P_N\approx0.281$ for $\beta^{-1}/m=100$.}
\end{figure}

Finally, we demonstrate the probability distribution of the number of produced pairs, $P_N\equiv P(N,N)$, by performing the integrations in Eqs.~(\ref{eq:characteristic}), (\ref{eq:unresolved}) numerically.
To this end, we specify the Bogoliubov coefficient $|\gamma_{s,\p}|^2$ to be the form of Eq.~(\ref{eq:coefficient}) with the transient effects neglected.
We also set $qE_0=m^2/2$ and $T=V^{1/3}=50/m$, so that the mean number of produced pairs in the vacuum corresponds to
\begin{align}
\frac{\<N\>_0}{2S+1} = \frac{(qE_0)^2TV}{(2\pi)^3}\exp\!\left(-\frac{\pi m^2}{qE_0}\right)
\approx 11.8.
\end{align}
The resulting $P_N$ for scalars ($S=0$) and spinors ($S=1/2$) are shown in Figs.~\ref{fig:scalar} and \ref{fig:spinor}, respectively, at three different temperatures and vanishing chemical potential.
We indeed find that the pair production is enhanced (suppressed) for scalars (spinors) with increasing the temperature due to the Bose stimulation (Pauli blocking)~\cite{Kim:2009}.
Furthermore, the probability distribution allows us to extract much more information than just the mean value.
For example, it is evident that the event-by-event fluctuation becomes more significant for scalars at higher temperatures and the pair annihilation becomes more probable for spinors as a consequence of the suppressed pair production.

We also note that, because of $|\gamma_{s,\p}|^2\leq\exp[-\pi m^2/(qE_0)]\approx0.00187\ll1$ in our parameter choice, the probability distributions shown in Figs.~\ref{fig:scalar} and \ref{fig:spinor} are practically indistinguishable from the bidirectional Poisson distribution in Eq.~(\ref{eq:bidirectional}).
Its two parameters in the low-temperature limit $\beta\to\infty$ are 
\begin{align}
X \to \<N\>_0 + \O(e^{-\beta m\pm\beta\mu}), \qquad
Y \to \O(e^{-2\beta m})
\end{align}
both for scalars and for spinors, whereas those in the high-temperature limit $\beta\to0$ are
\begin{align}
X-Y \to \O(\beta^{-1}), \qquad
X+Y \to \O(\beta^{-2})
\end{align}
for scalars but
\begin{align}
X-Y \to \O(\beta), \qquad
X+Y \to \frac{\<N\>_0}{2} + \O(\beta^2)
\end{align}
for spinors.
Recalling that the mean value and the standard deviation are provided by $X-Y$ and $\sqrt{X+Y}$, respectively, the numerical results are qualitatively understandable from the above asymptotic behaviors.

\section{Summary}
In summary, we studied the probability distribution of the number of particle and antiparticle pairs produced via the Schwinger effect when a uniform but time-dependent electric field is applied to noninteracting scalars or spinors initially at a thermodynamic equilibrium.
The central outcome of this paper is the formula for the characteristic function presented in Eq.~(\ref{eq:characteristic}), from which the probability distributions as well as their cumulants can be extracted in various situations.
In particular, when the numbers of produced pairs are fully counted per spin and momentum, the probability distribution obeys the generalized trinomial statistics [Eq.~(\ref{eq:trinomial})].
On the other hand, when only the total number of produced pairs is counted without resolving their spins and momenta, the probability distribution obeys the bidirectional Poisson statistics under an approximation reasonable to the Schwinger effect [Eq.~(\ref{eq:bidirectional})].

Physically, we found that the Schwinger pair production in a medium is enhanced for scalars due to the Bose stimulation but suppressed for spinors due to the Pauli blocking.
Furthermore, in addition to the production of accelerated pairs by the electric field, the annihilation of decelerated pairs was found to take place in a medium.
We hereby propose to refer to the latter phenomenon as ``Schwinger pair annihilation.''
The Schwinger pair annihilation becomes more probable for spinors at higher temperatures as a consequence of the suppressed pair production (Fig.~\ref{fig:spinor}).
We also found that the event-by-event fluctuation becomes more significant for scalars at higher temperatures (Fig.~\ref{fig:scalar}).

Because only the Bogoliubov coefficients are needed as input parameters for Eq.~(\ref{eq:characteristic}), it is applicable to a variety of time-dependent electric fields such as shaped laser pulses~\cite{Dumlu:2010,Dumlu:2011} and pulse sequences~\cite{Akkermans:2012}.
Hopefully, by linking the Schwinger pair production and annihilation in high-energy physics with the full counting statistics in mesoscopic physics, our formula may provide useful insights into future experimental data from high-intensity laser facilities as well as from relativistic heavy-ion collisions.
In particular, the latter are capable of creating high-temperature media at the same time as strong electric fields in principle.

\acknowledgments
The author thanks Koichi Hattori for valuable discussions.
This work was supported by JSPS KAKENHI Grants No.\ JP18H05405 and No.\ JP21K03384.


\begin{thebibliography}{99}
\bibitem{footnote2}
The bidirectional Poisson distribution is also known as the Skellam distribution in probability theory.
\end{thebibliography}
\end{document}